\documentclass[preprint,preprintnumbers,amsmath,amssymb]{revtex4}
\usepackage{graphicx}
\usepackage{graphics}
\usepackage{dcolumn}
\usepackage{bm}
\usepackage{amsfonts}
\usepackage{indentfirst}
\usepackage{setspace,lscape}
\usepackage{longtable}

 \setlongtables

\begin{document}

%\preprint{APS/123-QED}

\title{Relativistic effects on information measures for hydrogen-like atoms}

\author{K.~D.~Sen}\email{sensc@uohyd.ernet.in}
\affiliation{School of Chemistry, University of Hyderabad,
Hyderabad 500046, India}

\author{Jacob Katriel}\email{jkatriel@tx.technion.ac.il}
\affiliation{Department of Chemistry, Technion, Haifa 32000,
Israel}

\date{\today}

\begin{abstract}

Position and momentum information measures are evaluated for the
ground state of the \emph{relativistic} hydrogen-like atoms.
Consequences of the fact that the radial momentum operator is not
self-adjoint are explicitly studied, exhibiting fundamental
shortcomings of the conventional uncertainty measures in terms of
the radial position and momentum variances. The Shannon and
R\'enyi entropies, the Fisher information measure, as well as
several related information measures, are considered as viable
alternatives. Detailed results on the onset of relativistic
effects for low nuclear charges, and on the extreme relativistic
limit, are presented. The relativistic position density decays
exponentially at large $r$, but is singular at the origin.
Correspondingly, the momentum density decays as an inverse power
of $p$. Both features yield divergent R\'enyi entropies away from
a finite vicinity of the Shannon entropy. While the position space
information measures can be evaluated analytically for both the
nonrelativistic and the relativistic hydrogen atom, this is not
the case for the relativistic momentum space. Some of the results
allow interesting insight into the significance of recently
evaluated Dirac-Fock vs. Hartree-Fock complexity measures for
many-electron neutral atoms.

\end{abstract}

\maketitle


\noindent keywords: relativistic H atom; Heisenberg uncertainty
relation; Shannon and R\'enyi entropies; Fisher information measure.

\vfill\eject

\section{Introduction}

The celebrated Heisenberg uncertainty principle \cite{Heis1,Ken1}
of quantum mechanics is specified by means of the position and
momentum variances, that are defined in terms of the expectation
values of the corresponding (hermitian) operators. Several authors
have pointed out that for bimodal distributions this formulation
does not provide an adequate measure of the uncertainty of the
measurable involved. A variety of information measures have been
proposed and investigated in a fairly broad range of contexts. The
most familiar of these are due to Shannon \cite{Shannon48} and
Fisher \cite{Fisher25}. They are being increasingly applied in
studying the electronic structure and properties of atoms and
molecules, and play an important role in the rapidly developing
field of quantum information and its anticipated technological
offspring.

Due to its fundamental importance in natural sciences, the
hydrogen atom has been extensively studied from the
\emph{information theoretical} view point. Consequently, the
Shannon entropies \cite{Gadre100,Yanez102,Sen01} and Fisher
information \cite{Dehesa06} of the non-relativistic hydrogen-like
atom have been studied in considerable detail. We note, with some
surprise, a glaring omission, the relativistic hydrogen atom,
formulated in terms of the Dirac equation, which is the subject of
the present study. Future studies of relativistic effects on the
information measures in many-electron atoms will certainly benefit
from the presently derived results. Indeed, very recent work on
Dirac-Fock vs. Hartree-Fock complexity measures for neutral
many-electron atoms \cite{Borgoo} allows certain comparisons to be
made with results obtained in the present article, that shed
additional light on the significance of that study.

The Shannon information entropy $S_{r}$ of
the spatial electron density $\rho(\textbf{r})$ is
defined as
\begin{equation}\label{eq:eq1}
    S_{r}=-\int
    \rho(\textbf{r})\,\ln{\rho(\textbf{r})}\,d\textbf{r} \; ,
\end{equation}
and the corresponding momentum space entropy $S_{p}$ is given by
\begin{equation}\label{eq:eq2}
    S_{p}=-\int \Pi(\textbf{p})\,\ln{\Pi(\textbf{p})}\,d\textbf{p} \; ,
\end{equation}
where $\Pi(\textbf{p})$ denotes the momentum density. The
densities $\rho(\textbf{r})$ and $\Pi(\textbf{p})$ are each
normalized to unity and all quantities are given in atomic units.
These two densities are obtained from the corresponding position
and momentum space wavefunctions, that are the Fourier transforms
of one another. The Shannon entropy sum $S_{T}=S_{r}+S_{p}$
contains the net information and obeys the well known lower bound
derived by Bialynicki-Birula and Mycielski \cite{Bialynicki75},
\begin{equation}\label{eq:eq3}
    S_{T}=S_{r}+S_{p} \geq n\,(1+\ln{\pi}) \; ,
\end{equation}
where $n$ is the number of dimensions. The lower bound is attained
by a Gaussian distribution. This entropic uncertainty-like
relation represents a stronger version of the Heisenberg
uncertainty principle of quantum mechanics. The individual
entropies $S_{r}$ and $S_{p}$ depend on the units used to measure
$r$ and $p$ respectively, but their sum $S_{T}$ does not, i.e., it
is invariant under uniform scaling of coordinates.

The Shannon entropies provide a global
measure of information about the probability distribution in the
respective spaces. A more localized distribution
yields a \emph{smaller} value of the corresponding information entropy. For
applications of Shannon information entropy in chemical physics we
refer the reader to the published literature
\cite{Gadre100,Smith101}.

In the context of the quantum theory of one-particle systems the
Fisher position information measure is defined as
\begin{equation}\label{eq:eq4}
    I_{r}=\int \frac{\left|\nabla\rho(\textbf{r})\right|^2}{\rho(\textbf{r})}
    \,d\textbf{r}
\end{equation}
and the corresponding momentum space measure is given by
\begin{equation}\label{eq:eq5}
    I_{p}=\int \frac{\left|\nabla \Pi(\textbf{p})\right|^2}{\Pi(\textbf{p})}
    \,d\textbf{p}.
\end{equation}
For a general definition of the Fisher information measure and a
careful exposition of its significance we refer to the definitive
monograph by Rao \cite{Rao59}.

The individual Fisher measures are bounded through the Cramer-Rao
inequality \cite{Rao59,Stam1} according to
$\displaystyle{I_{r}\geq \frac{1}{V_{r}}}$ and
$\displaystyle{I_{p}\geq \frac{1}{V_{p}}}$, where $V$'s denote the
corresponding spatial and momentum variances, respectively. In
position space the Fisher information measures the sharpness of
the probability density, and for a Gaussian distribution is
exactly equal to the inverse of the variance \cite{Frieden04}. A
sharp (smooth) and strongly localized (well spread-out)
probability density gives rise to a \emph{larger} (\emph{smaller})
value of the Fisher information in the position space. With a
differential probability density as its content, the Fisher
measure is better suited to study the localization characteristics
of the probability distribution than the Shannon information
entropy \cite{Carroll06,Garba01}. Unlike $S_r+S_p$, for which eq.
\ref{eq:eq3} specifies a lower bound, general bounds are as yet
unknown for the Fisher product $I_{r}I_{p}$. Since localization
(i.e., low uncertainty) means high values of the Fisher
information measures, the counterpart of the Heisenberg or Shannon
bound should be an upper bound on the product of position\ and
momentum Fisher information measures. For a single particle under
the influence of a central potential Dehesa {\it{et al.}}
\cite{Dehesa011} have very recently reported a lower bound on the
Heisenberg product \cite{Heis2,Heis3,Heis4,Heis5} which can be
directly related to the Fisher information. For the application of
the Fisher information measure as an underlying guideline for the
formulation of fundamental physical principles we refer to the
recent book by Frieden \cite{Frieden04}, and for applications to
the electronic structure of atoms, to the pioneering work of
Dehesa {\it{et al.}} \cite{Dehesa01,Dehesa02,Dehesa03,Dehesa04}.

A widely used generalization of the Shannon entropy is the R\'enyi
entropy. The R\'enyi position entropy (that, when a more precise
designation is required, we shall address as the $a$-R\'enyi
position entropy) is defined as \cite{IBB}
\begin{equation}
\label{eq:Re} H_{a}^{(r)}=\frac{1}{1-a}\log\left(
      \int_0^{\infty}[\rho(r)]^{a} 4\pi r^2 dr \right) \; .
\end{equation}
The symmetrized R\'enyi position entropy is
\begin{equation}
\label{eq:Hs} {\cal{H}}_s^{(r)}=(H_{a}^{(r)}+H_{b}^{(r)})/2
\end{equation}
where
$$a=\frac{1}{1-s}, \;\;\;\; b=\frac{1}{1+s},
   \;\;\;\; -1\leq s \leq 1, \;\;\; {\mbox{i.e.,}} \;
     \frac{1}{2}\leq a,b \leq \infty, \;\;\;
     \frac{1}{a}+\frac{1}{b}=2 .$$
The R\'enyi momentum entropies are similarly defined in terms of
the momentum density $\Pi(p)$. For $a=1$ one obtains, using
l'H\^opital's rule,
$$\lim_{a\rightarrow 1} H_{a}^{(r)} =
  -\int_0^{\infty} \rho(r) \log[\rho(r)] 4\pi r^2 dr = S_r \; ,$$
where $S_r$ is the Shannon position entropy \cite{Shannon48}. For
an $n$ dimensional system, the sum of the $a$-R\'enyi position
entropy and the $b$-R\'enyi momentum entropy, for $a$ and $b$
satisfying $\frac{1}{a}+\frac{1}{b}=2$, was recently shown by
Bialynicki-Birula \cite{IBB} to satisfy the inequality
(uncertainty-like relation)
\begin{equation}
\label{eq:RUR} H_a^{(x)}+H_b^{(p)} \geq
n\left[\frac{1}{2}\left(\frac{\log(a)}{a-1}
+\frac{\log(b)}{b-1}\right)+\log(\pi)\right] \; .
\end{equation}

The properties of the Shannon entropies and several other
information measures under coordinate scaling have been examined
in ref. \cite{SK}. It was pointed out that upon scaling the
coordinates via $\tilde{r}=\zeta r$, normalization of the density
requires that
$$\tilde{\rho}(r)
        =\frac{1}{\zeta^3}\rho\left(\frac{r}{\zeta}\right) \; .$$
It follows that
\begin{eqnarray*}
\tilde{H}_{a}^{(r)} &=& \frac{1}{1-a}\log\left(
\int_0^{\infty}[\tilde{\rho}(r)]^{a} 4\pi r^2 dr \right) \\
&=& \frac{1}{1-a}\log\left(\zeta^{3(1-a)}
\int_0^{\infty}[\rho(x)]^{a} 4\pi x^2 dx \right)
=3\log(\zeta)+H_{a}^{(r)} \; .
\end{eqnarray*}

The scaling of the coordinates introduced above implies scaling of
the momenta according to $\tilde{p}=\frac{p}{\zeta}$, so
$$\tilde{H}_{a'}^{(p)}=-3\log(\zeta)+H_{a'}^{(p)} \; .$$
Hence,
$$\tilde{H}_{a}^{(r)}+\tilde{H}_{a'}^{(p)}
    =H_{a}^{(r)}+H_{a'}^{(p)} \; .$$
Note that $a$ and $a'$ are entirely independent of one another.

For a system whose hamiltonian is of the form $
{\cal{H}}=\hat{T}+\lambda V(r)\, ,$ where
$\hat{T}=-\frac{1}{2}\nabla^2$, if the potential is homogeneous,
i.e., $V(\zeta r)=\zeta^k V(r)$, scaling of the coordinates via
$\tilde{r}=\zeta r$ is equivalent to scaling the coupling constant
via $\tilde{\lambda}=\zeta^{k+2}\lambda$ \cite{SK}. Hence, for
such potentials the sum of an $a$-R\'enyi position entropy and an
$a'$-R\'enyi momentum entropy is independent of the coupling
constant $\lambda$.

Since the position density has dimensions of inverse volume and
the momentum density has dimensions of inverse momentum cubed, the
Shannon, Fisher and R\'enyi entropies can be converted into
quantities that have dimensions of length or momentum. This
property is used to facilitate comparison among the different
information measures. We refer to these transformed quantities as
``length'' and ``impetus'', respectively. This practice allows a
clear distinction between variance-based uncertainty measures and
information based measures that have units of position or momentum
although they do not involve expectation values of the
corresponding quantum-mechanical operators. ``Impetus'' is a
pre-Newtonian synonym of momentum.

A different class of information measures, involving ratios
between relativistic and nonrelativistic densities
\cite{Kul01,Raju01}, is considered as well. Since these
information measures are pure numbers they cannot be transformed
into quantities with units of length or momentum. However, they
allow the onset of relativistic effects upon increase of the
nuclear charge to be followed very transparently.

A third class of information measures is represented by the
Tsallis entropy \cite{Tsalis01},
$$\epsilon_a=
\frac{1}{a-1}\left(1-4\pi\int_0^{\infty}(\rho(r))^a r^2dr\right)\;
,$$ that has been invoked for non-extensive systems, and which is
not homogeneous under coordinate scaling. The Tsallis entropy is
closely related to the R\'enyi entropy via
$$\epsilon_a=\frac{1}{a-1}\left(1- \exp\big((1-a)H_a^{(r)} \big) \right) \; .$$
For $a\rightarrow 1$ these two entropies coincide with one another
as well as with the Shannon entropy.

Upon attempting to evaluate the various uncertainty and
information measures for the relativistic hydrogen atom we
encountered three somewhat surprising obstacles. The first has to
do with the fact that the radial momentum does not have a proper
(self-adjoint) quantum-mechanical counterpart. This fact has been
known for a while, and its consequences in the present context are
explained below. Another surprise had to do with the fact that the
momentum-space solutions of the Dirac equations are still subject
to controversy \cite{Dirac,Dirac1}. Finally, we find it intriguing
that for the relativistic hydrogen atom the position space
information measures could be evaluated analytically, but for the
momentum space measures we had to apply numerical integration. The
first difficulty suggests that the various uncertainty-like
principles that do not involve the variance of position or
momentum have even stronger merits for multi-dimensional systems
than those extensively pointed out in previous studies on one
dimensional systems. We addressed the second difficulty by
adhering to the version of the momentum wavefunction that in our
judgement was the most natural and straightforward \cite{Sheth01}
without bothering to explore its equivalence (or lack of it) with
other formulations.

This article is structured as follows: In sections 2 and 3 we
consider the radial position and momentum variances for the
nonrelativistic and the relativistic hydrogen atom, respectively,
presenting very explicitly the consequences of the non
self-adjointness of the commonly invoked radial momentum operator.
The Shannon entropies are investigated in section 4, and the
Fisher information measures in section 5. In section 6 we study
the R\'enyi entropies, also considering the average density, that
is essentially a special case of these. This allows a brief
discussion of complexity measures, with a rather surprising
comparison with a recent Dirac-Fock study of neutral many-electron
atoms \cite{Borgoo}. Scale invariant entropies are discussed in
section 8, and some concluding remarks are made in section 9.

\section{Uncertainty measures for the nonrelativistic \hfill\break
hydrogen-like atoms}

The most familiar measures of uncertainty are the position and
momentum variances, which for a one dimensional system defined
over the whole real axis are given by \hfill\break $(\delta
x)=\sqrt{<x^2>-<x>^2}$ and $(\delta p)=\sqrt{<p^2>-<p>^2}$, where
\begin{eqnarray}
\label{eq:xp}
<x^k> &=& \int_{-\infty}^{\infty} x^k|\psi(x)|^2 dx =
      \int_{-\infty}^{\infty}\phi^*(p)
               \left(i\hbar\frac{d}{dp}\right)^k\phi(p) dp \\
<p^k> &=& \int_{-\infty}^{\infty} p^k|\phi|^2 dp =
      \int_{-\infty}^{\infty}\psi^*(x)
               \left(-i\hbar\frac{d}{dx}\right)^k\psi(x) dp \; .\nonumber
\end{eqnarray}
Here, $\psi(x)$ and $\phi(p)$ are the position and momentum
wavefunctions, which are the Fourier transforms of one another.
Eq. \ref{eq:xp} emphasizes the correspondence between position and
momentum space expectation values of the (hermitian) position and
momentum operators, a correspondence that, as we shall explicitly
demonstrate below, fails for the (non self-adjoint) radial
momentum operator.

\subsection{Position and momentum variances for spherically symmetric
three-dimensional systems}

Qiang and Dong \cite{Qiang}, following a time-honored tradition, suggest
that the radial momentum operator in the coordinate
representation, using atomic units in which $\hbar=1$, is
$$p_r = -i\left(\frac{\partial}{\partial r} + \frac{1}{r} \right) \; .$$
They justify this expression by noting that
$$p_r^2 = -\left( \frac{\partial^2}{\partial r^2}
                    + \frac{2}{r}\frac{\partial}{\partial r} \right) $$
is (up to a multiplicative constant) the radial part of the
Laplacian (i.e., the kinetic energy operator).

Note, however, that the $n$ dimensional generalization
$$p_r = -i\left(\frac{\partial}{\partial r} + \frac{n-1}{2 r}\right) $$
satisfies
$$ -p_r^2 = \frac{\partial^2}{\partial r^2}
     + \frac{n-1}{r} \frac{\partial}{\partial r}
     + \frac{(n-1)(n-3)}{4r^2} \; , $$
which, for $n=2$ and $n>3$ does not agree with the radial part of
the Laplacian,
$${\cal{L}}_r = \frac{\partial^2}{\partial r^2}
                     + \frac{n-1}{r} \frac{\partial}{\partial r} \, .$$

Straightforward integration by parts yields
\begin{equation} \label{eq:pr}
<\psi|p_r|\psi> = -i S_n\int_0^{\infty} \psi(r)
\left[\left(\frac{\partial}{\partial r}+\frac{n-1}{2
r}\right)\psi(r)\right] r^{n-1}dr = 0
\end{equation}
where $S_n=\frac{2
\pi^{\frac{n}{2}}}{\Gamma\left(\frac{n}{2}\right)}$ is the surface
area of the $n$-dimensional unit sphere.

Paz \cite{Paz2} has recently rigorously shown that $p_r$ is not
self-adjoint, and has no self-adjoint extension ({\it{cf.}}, also,
\cite{Liboff1,Liboff2,Liboff3}, for earlier discussions of this
issue). This fact, which is more closely considered below,
suggests that using the variance of the radial momentum as an
uncertainty measure for (spherically symmetric) three-dimensional
systems may be questionable.

\subsection{Position space wavefunction of the nonrelativistic hydrogen atom}

The ground state nonrelativistic hydrogenic wavefunction
$\psi_{0,0,0}(r) = \left(\frac{Z^3}{\pi}\right)^{\frac{1}{2}}
\exp(-Z r)$ yields the density
\begin{equation}
\label{eq:rhoNR}
\rho_{NR}(r) =  \frac{Z^3}{\pi} \exp(-2Z r) \, .
\end{equation}
The expectation values $<r> = \frac{3}{2 Z}$ and $<r^2> =
\frac{3}{Z^2} $ yield
$$(\delta r) = \sqrt{<r^2>-<r>^2}
       = \frac{\sqrt{3}}{2 Z} \approx \frac{0.8660}{Z} \; .$$
Similarly, $<p_r>=0$ and $<p_r^2> = Z^2$.
The latter value is equal to $<-\nabla^2>$, because, for a spherically
symmetric wavefunction, the angular part of the
Laplacian makes a vanishing contribution.

Using these position space expectation values
to evaluate $(\delta p_r)=Z$, we obtain the uncertainty product
$(\delta r)(\delta p_r) \approx 0.8660$.

\subsection{Momentum space wavefunction of the nonrelativistic hydrogen atom }

The momentum space wavefunction for the ground state of the
nonrelativistic hydrogen atom \cite{MF2}
\begin{equation}
\label{eq:eq6} \chi_{0,0,0}(p) = \frac{1}{\pi}
\left(\frac{2}{Z}\right)^{\frac{3}{2}}
                      \left[ \left(\frac{p}{Z}\right)^2 + 1\right]^{-2} \; ,
\end{equation}
yields the momentum density
\begin{equation}
\label{eq:eq8} \Pi_{NR}(p)=\frac{8}{\pi^2Z^3}\left[\left(
\frac{p}{Z}\right)^2+1\right]^{-4}\; ,
\end{equation}
in terms of which we obtain
$$<p> = \int_0^{\infty} p \Pi_{NR}(p) 4\pi p^2 dp =
                  \frac{8Z}{3\pi} \approx 0.848826 Z \; ,$$
and
$$<p^2> = \int_0^{\infty} p^2 \Pi_{NR}(p) 4\pi p^2 dp = Z^2 \; . $$
The latter value agrees with the position space expectation value
of $\; -\nabla^2$, but the former does not agree with the position
space result, eq. \ref{eq:pr}.

Using these momentum space expectation values we obtain
$$(\delta p) = \frac{Z}{3\pi}\sqrt{9\pi^2-64} \approx 0.5287 Z \; .$$
Along with the value of $\delta r$ obtained above we get
$$(\delta r)(\delta p) \approx 0.4578 \; ,$$
which is less than $\frac{1}{2}$. This is probably a manifestation of
the questionable status of the radial momentum, pointed out above.

Messiah \cite{Messiah} shows that, in one dimension, $<x^2><p_x^2>
\; \geq \; \frac{1}{4}$. For a spherically symmetric system
$<r^2>=3<x^2>$ and $<p^2>=3<p_x^2>$. Hence, $<r^2><p^2> \; \geq
\frac{9}{4}$. The results quoted above imply that for the
nonrelativistic hydrogen atom $<r^2><p^2>=3$, which is larger than the
lower bound derived by Messiah.

We conclude this section by emphasizing that the operators $r^2$
and $p^2$ can be expressed (in Cartesian coordinates) in terms of
manifestly self-adjoint operators. This is not the case for $r$
and $p$. Further consequences of this distinction are presented in
the following section.

\section{Uncertainty measures for the relativistic hydrogen-like atoms}

\subsection{Relativistic position uncertainty}

For the ground state of the Dirac hydrogenic atom the (spin up)
wavefunction is of the form
\begin{equation}
\label{eq:eq10} \Psi_D = N r^{\gamma-1} \exp(-Z r)\left(
  \begin{array}{l}
    G \\
    0 \\
    igY_0 \\
    igY_1
  \end{array} \right)
\end{equation}
where $G=\sqrt{1+\gamma}$, $g=\sqrt{1-\gamma}$, $Y_0=\cos(\theta)$ and
$Y_1=\sin(\theta)\exp(i\phi)$.
The normalization factor is given by
$N=\frac{(2Z)^{\gamma+\frac{1}{2}}}{\sqrt{8\pi\Gamma(2\gamma+1)}} $
where
$\gamma=[ 1-(Z\alpha)^2 ]^{\frac{1}{2}}$.
Here $\alpha \approx \frac{1}{137.03600}$ is the fine-structure constant.
In the limit $Z\rightarrow 0$
or $c \rightarrow \infty$ [remembering that in atomic units $\alpha=1/c$]
it follows that $\gamma \rightarrow 1$
and we obtain the nonrelativistic wavefunction
$\sqrt{\frac{Z^3}{\pi} } \exp(-Zr)$.

The ground state position wavefunction yields the position density
\begin{equation}
\label{eq:eq11}
\rho_{R}(r) = \frac{(2 Z)^{2\gamma+1}}{4 \pi
\Gamma(2\gamma+1)}
                  r^{2(\gamma-1)} \exp(-2 Z r) \; .
                  \end{equation}
In the extreme relativistic limit $Z\rightarrow\frac{1}{\alpha}$
the position density obtains the form \cite{Garcia}
$$\rho_{ER}(r) = \frac{1}{2\pi\alpha} r^{-2}
     \exp\left(-\frac{2r}{\alpha}\right) \, .$$

Using the relativistic wavefunction $\Psi_D$, eq. (\ref{eq:eq10}),
we obtain
$$<r>_R=\frac{2\gamma+1}{2Z}$$
and
$$<r^2>_R=\frac{(\gamma+1)(2\gamma+1)}{2Z^2}
  \approx <r^2>_{NR} - \frac{1}{Z^2}\left(\frac{7}{4}(\alpha Z)^2
     +\frac{3}{16}(\alpha Z)^4 +\cdots \right) \; .$$
hence,
$$(\delta r)=\frac{\sqrt{2\gamma+1}}{2 Z} \; .$$
For future reference we define the ratio
%$\lambda\equiv\sqrt{\frac{<r^2>_R}{<r^2>_{NR}}}$, which is plotted
$\beta\equiv\sqrt{\frac{<r^2>_R}{<r^2>_{NR}}}$, which is plotted
in Fig. 1. Evaluating $<p_r>_R = 0$ and
\begin{equation}
\label{eq:eq12}
 <p_r^2>_R = \frac{Z^2}{2\gamma-1} \approx <p_r^2>_{NR}
      +Z^2\left( (\alpha Z)^2 +\frac{5}{4}(\alpha Z)^4 +\cdots\right)
\end{equation}
it follows that $(\delta p_r)=\frac{Z}{\sqrt{2\gamma-1}}$.
Hence,
$$(\delta r)(\delta p_r) = \frac{1}{2}\sqrt{\frac{2\gamma+1}{2\gamma-1}} \; .$$

A singularity is observed for $2\gamma-1=0$, that yields
$Z=\frac{\sqrt{3}}{2\alpha} \approx 118.68$.
We do not know what significance to assign to this nuclear charge.

Since the small components of the Dirac wavefunction for the
hydrogen atom depend on the angular coordinates, the expectation value
of the Laplacian is not the same as that of ${\cal{L}}_r$, in spite of the
fact that the ground state density is spherically symmetric.
The angular part of the Laplacian yields
\begin{eqnarray*}
& & <\Psi_D|\frac{1}{r^2}\hat{L}^2|\Psi_D> = \\
& & {\phantom{leave s}}
                 = \int_0^{\infty} N^2 r^{2(\gamma-1)} \exp(-2 Z r) dr
\left<\Big(G, \; 0,\; -igY_0, \;
   -ig{\overline{Y_1}} \big) \big{|} \hat{L}^2 \big{|}
    \left(\begin{array}{l}
                 G \\
                 0 \\
                 igY_0 \\
                 ig Y_1 \end{array}
                              \right)\right> = \nonumber   \\
& & {\phantom{leave more space}} =
4Z^2\frac{1-\gamma}{2\gamma(2\gamma-1)} \; .
\end{eqnarray*}
Adding the value of $<\Psi_D|p_r^2|\Psi_D>$, eq. \ref{eq:eq12}, we obtain
\begin{equation}
\label{eq:eq14} <\Psi_D|-\nabla^2|\Psi_D> =
\frac{2-\gamma}{\gamma(2\gamma-1)} Z^2 \; .
\end{equation}
The singularity at $Z\approx 118.68$ remains

\subsection{Relativistic momentum uncertainty}

We use the expression for the relativistic ground state momentum
wavefunction due to Sheth \cite{Sheth01}.
Denoting the radial momentum
variable by $p$ and defining
\begin{equation}
\label{eq:x}
x=(\gamma+1) \arctan(\frac{p}{Z}) \; ,
\end{equation}
the momentum density can be written in the form
$$\Pi_R(p)= \frac{\Gamma(\gamma+1)}
                {2 Z^3 \pi^{\frac{3}{2}}\Gamma(\gamma+\frac{1}{2})}
          \frac{F(p)}{(\frac{p}{Z})^2\Big((\frac{p}{Z})^2+1\Big)^{\gamma+1}}$$
where
\begin{equation}
\label{eq:F}
F(p)=(\gamma+1) \sin^2 (x) +
                \frac{1-\gamma}{\gamma^2} R(p)^2
\end{equation}
and $R(p)=(\gamma+1) \cos(x) - \frac{Z}{p} \sin(x)$.
$\Gamma$ is the familiar $\Gamma$-function.

In the nonrelativistic limit $\gamma\rightarrow 1$ this expression
reduces to $\Pi_{NR}(p)$, {\it{cf.}} eq. (\ref{eq:eq8}).

In the extreme relativistic limit, $Z \rightarrow \frac{1}{\alpha}$, the
momentum density is obtained by using L'H\^opital's rule to evaluate
$\lim_{\gamma\rightarrow 0} \frac{R}{\gamma}
       =\frac{\partial R}{\partial \gamma}\Big{|}_{\gamma=0}$.
It is found to be of the form
\begin{equation}
\label{eq:eq15} \Pi_{ER}(p) = \frac{\alpha^3}{2\pi^2}
\frac{1}{\left[\alpha p(1+(\alpha p)^2)\right]^2}
    \left[ (\alpha p)^2
   + \left(1-( \alpha p+\frac{1}{\alpha p})
              \arctan(\alpha p)\right)^2\right] \, .
\end{equation}

The long range decay of the momentum density can be established by
noting that for $p\rightarrow\infty$
$$\frac{1}{\left(\frac{p}{Z}\right)^2
       \left(\left(\frac{p}{Z}\right)^2+1\right)^{\gamma+1}}
         \approx \left(\frac{Z}{p}\right)^{(2\gamma+4)} \; .$$
Furthermore,
$$x \approx (\gamma+1)\frac{\pi}{2} \; ,$$
so that
$$F\approx F_{\infty} \equiv (\gamma+1) \left( 1 +
  \frac{1-2\gamma^2}{\gamma^2}
          \cos^2\left((\gamma+1)\frac{\pi}{2}\right)\right) \; .$$
To evaluate $<p^2>$ we integrate over $p$ numerically between
$p=0$ and $p=p_m$, where $p_m$ is chosen large enough for the density to
be close enough to its asymptotic form, and add an integral over the
asymptotic momentum density between $p=p_m$ and $p=\infty$. Hence,
$$<p^2>_R = 4\pi \int_0^{p_m} \Pi_R(p) p^4 dp +
          \frac{2\Gamma(\gamma+1)}{\sqrt{\pi}\Gamma(\gamma+\frac{1}{2})}
             Z^{2\gamma+1} F_{\infty} \frac{p_m^{-(2\gamma-1)}}{2\gamma-1} \; . $$
%The ratio $\phi\equiv\sqrt{\frac{<p^2>_R}{<p^2>_{NR}}}$ is plotted
The ratio $\mu\equiv\sqrt{\frac{<p^2>_R}{<p^2>_{NR}}}$ is plotted
in Fig. 1. The expectation value $<p^2>_R$ was evaluated for the
values of $Z$ considered by Qiang and Dong \cite{Qiang}. The
results agree with those evaluated analytically by using the
position space expectation value of the Laplacian, eq.
(\ref{eq:eq14}), for $Z=1, \; 11, \; 37$, to ten decimal places,
and for $Z=87$ to eight decimal places. The values of
$\sqrt{<p_r^2>}$, evaluated in terms of the ground-state
wavefunction in the position representation, using Eq.
(\ref{eq:eq12}) , agree with the values of $(\Delta p_r)_R$ in
\cite{Qiang}. It would be nice to show analytically that $<p^2>$
is indeed equal to the right hand side of Eq. (\ref{eq:eq14}).

The results presented above clearly expose the difference between
the self-adjoint Laplacian and the non-self adjoint radial
momentum.

\section{Shannon entropies for the hydrogen atom}

\subsection{Nonrelativistic position and momentum entropies}

The nonrelativistic position entropy $S_r^{NR}=-\int_0^{\infty}
4\pi r^2\rho_{NR}(r)\log\Big(\rho_{NR}(r)\Big)dr$ can be easily evaluated
for the hydrogenic nonrelativistic ground state position density, yielding
\begin{equation}
\label{eq:eq7}
S_r^{NR} = 3+\log(\pi)-3\log(Z) \approx 4.1447299 -3\log(Z) \; .
\end{equation}
Similarly, the nonrelativistic momentum entropy
$S_p^{NR}=\int_0^{\infty} 4\pi p^2\Pi_{NR}(p)\log\Big(\Pi_{NR}(p)\Big) dp$
can be evaluated for the hydrogenic nonrelativistic ground state
momentum density, yielding
\begin{equation}\label{eq:eq9}
S_p^{NR} = -\frac{10}{3}+5\log(2)+2\log(\pi)+3\log(Z)
       \approx 2.4218623 + 3\log(Z) \, .
       \end{equation}
It follows that
\begin{eqnarray*}
S_r^{NR} + S_p^{NR} &=& -\frac{1}{3} + 5\log(2) +3\log(\pi)
       \approx 6.5665922 \\
                    &>& 3(1+\log(\pi)) \approx 6.4341897 \; .
\end{eqnarray*}

\subsection{Shannon length and impetus}

Since $\exp(S_r)$ has dimensions of volume, we define the Shannon length
$R_S$ via
$$\frac{4\pi}{3} R_S^3 = \exp(S_r) \, .$$
We similarly define the Shannon impetus $P_S$ via
$$\frac{4\pi}{3} P_S^3 = \exp(S_p) \, .$$
For the nonrelativistic hydrogen atom we obtain
$$R_S^{NR} = \left(\frac{3}{4}\right)^{\frac{1}{3}}\frac{e}{Z}
             \approx \frac{2.46972}{Z}$$
and
$$P_S^{NR} = 2 (3\pi)^{\frac{1}{3}} \exp(-\frac{10}{9}) Z
             \approx 1.39071 Z \, ,$$
hence,
$$R_S^{NR} P_S^{NR} \approx 3.43466$$

From the entropic uncertainty-like relation
$$S_r+S_p \geq 3(1+\ln(\pi))$$
it follows that
$$R_S P_S \geq \left(\frac{3}{4\pi}\right)^{\frac{2}{3}}\pi e
   \approx 3.28639 \, .$$
This lower bound is, indeed, lower than the product of the
hydrogenic Shannon length and impetus, but fairly close to it.

\subsection{Relativistic position entropy}

The position entropy can be evaluated analytically in terms of the
relativistic ground state position density, eq. (\ref{eq:eq11}),
to yield
\begin{equation}
\label{eq:position}
 S_r^{R} = \log\left(\frac{\pi
\Gamma(2\gamma+1)}{2 Z^3}\right) + (2\gamma+1)
              -2(\gamma-1)\Psi(2\gamma+1)\,
\end{equation}
where $\Psi(z)$ is the Digamma function, defined as
$\Psi(z)=\frac{d\log(\Gamma(z))}{dz}$.

In the nonrelativistic limit $\gamma=1$ so
$\Gamma(2\gamma+1)=\Gamma(3)=2$ and $S_r^{R}$ reduces to
$S_r^{NR}$, eq. \ref{eq:eq7}.

The relativistic correction is
\begin{equation}
\label{eq:SrRmNR}
   S_r^{R} - S_r^{NR} = 2(\gamma-1)(1-\Psi(2\gamma+1))
    +\log\left(\frac{\Gamma(2\gamma+1)}{2}\right)
           \approx -(\alpha Z)^2
       +\left(\frac{3}{8}-\frac{\pi^2}{12}\right)(\alpha Z)^4 +\cdots \, .
\end{equation}

In the extreme relativistic limit
\begin{equation}
\label{eq:SER}
S_r=\log\left(\frac{\pi\alpha^3}{2}\right)-2C+1 \approx -14.463580
\end{equation}
where $C=0.5772156649..$ is Euler's constant (more commonly denoted
$\gamma$, a notation we avoid for an obvious reason).

\subsection{Relativistic momentum entropy}

The relativistic momentum entropy was evaluated by means of
numerical integration. To examine the relativistic corrections to
the position and momentum entropies more closely, we consider
$\sigma_r \equiv \frac{S_r^R-S_r^{NR}}{(\alpha Z)^2}$ and
$\sigma_p \equiv \frac{S_p^R-S_p^{NR}}{(\alpha Z)^2}$. Eq.
\ref{eq:SrRmNR} yields $\sigma_r \approx -1-0.4475(\alpha
Z)^2+\cdots$ and a numerical fit yields $\sigma_p \approx
1.80+0.65 (\alpha Z)^2+\cdots$, so $\sigma_r + \sigma_p \approx
0.80 +0.2 (\alpha Z)^2+\cdots$.

\subsection{Relativistic Shannon length and impetus}

For the relativistic Shannon length we obtain
$$R_S^R = \frac{1}{2Z} \Big(3\Gamma(2\gamma+1)\Big)^{\frac{1}{3}}
 \exp\left[1+\frac{2}{3}(\gamma-1)\Big(1-\Psi(2\gamma+1)\Big)\right] \, .$$
Hence,
%$$\lambda_S\equiv\frac{R_S^R}{R_S^{NR}} =
$$\beta_S\equiv\frac{R_S^R}{R_S^{NR}} =
    \left(\frac{\Gamma(2\gamma+1)}{2}\right)^{\frac{1}{3}}
    \exp\left[\frac{2}{3}(\gamma-1)\Big(1-\Psi(2\gamma+1)\Big)\right]
    \approx  1-\frac{1}{3}(\alpha Z)^2+\cdots \, .$$
The ratio of the relativistic and nonrelativistic Shannon
impeti can be obtained in terms of the numerically determined
$\sigma_p$, as follows,
%$$\phi_S\equiv\frac{P_S^R}{P_S^{NR}} =
$$\mu_S\equiv\frac{P_S^R}{P_S^{NR}} =
     \exp\left[\frac{(\alpha Z)^2}{3}\sigma_p\right] \approx
     1+0.60 (\alpha Z)^2 +  \cdots \, .$$
Hence,
\begin{equation}
\label{eq:RPS} \beta_S\mu_S \approx 1+0.27(\alpha Z)^2 +\cdots \,
.
\end{equation}

In Fig. 1 we present the ratios of the relativistic to
nonrelativistic Shannon lengths and impeti, along with the
corresponding ratios of the root mean square radius and momentum,
and that of the Fisher lengths and impeti (to be discussed below).
We note that for large $Z$ the relativistic effect on the momentum
uncertainty measures is larger than on their position
counterparts. This is most pronounced for the root mean square
position and momentum, whereas the relativistic effects on the
Fisher measures are almost symmetrical. This is most likely due to
the fact that the Fisher measures are sensitive to the local
oscillations of the distribution rather than to its long range
behavior, where the relativistic momentum distribution varies the
most.

\section{Fisher information measures for the hydrogen atom}

The Fisher position information measure, $I_r=\int_0^{\infty} 4\pi
r^2\frac{1}{\rho(r)} \left(\frac{d\rho}{dr}\right)^2 dr$ can be
easily evaluated for the ground state of the relativistic
hydrogen-like atom, yielding
$$I^R_r=\frac{4 Z^2}{2\gamma -1}=4<p_r^2> \approx
  I_r^{NR}\left(1+ (\alpha Z)^2+\frac{5}{4}(\alpha Z)^4
    + \frac{13}{8}(\alpha Z)^6 + \cdots\right) \; ,$$
where $I_r^{NR}=4 Z^2$.
Since $I_r^R$ has a singularity when $2\gamma-1=0$, {\it{i. e.,}} at
$Z \approx 118.68$, we do not examine the extreme relativistic limit.

The Fisher momentum information measure
$I_p=\int_0^{\infty} 4\pi p^2\frac{1}{\Pi(p)}
       \left(\frac{d\Pi}{dp}\right)^2 dp$
for the nonrelativistic momentum distribution can be evaluated
analytically, yielding $I^{NR}_p=\frac{12}{Z^2}$. For the
relativistic momentum density the Fisher information was evaluated
numerically.

\subsection{Fisher length and impetus}

The Fisher position information measure $I_r$ has dimensions of
inverse area. We define the Fisher length as
$R_F=I_r^{-\frac{1}{2}}$. Similarly, we define the Fisher impetus
$P_F=I_p^{-\frac{1}{2}}$. For the nonrelativistic hydrogen atom we
obtain $R_F^{NR} = \frac{1}{2Z}$ and $P_F^{NR} =
\frac{Z}{\sqrt{12}}$. The ratio between the relativistic and the
nonrelativistic Fisher lengths is
$$\beta_F\equiv\frac{R_F^R}{R_F^{NR}}=(2\gamma-1)^{\frac{1}{2}} \approx
    1-\frac{1}{2}(\alpha Z)^2-\frac{1}{4}(\alpha Z)^4+\cdots \, , $$
The corresponding ratio of Fisher impeti is evaluated by fitting
the numerically evaluated ratios to obtain
$$\mu_F\equiv\frac{P_F^{R}}{P_F^{NR}} \approx 1+0.4166(\alpha Z)^2+0.23 (\alpha Z)^4
      + \cdots \, , $$
hence,
\begin{equation}
\label{eq:RPF} \beta_F\mu_F \approx 1-0.0834(\alpha Z)^2+\cdots \,
.
\end{equation}
We note that the leading relativistic term is negative, unlike the
corresponding term for the product of Shannon length and impetus,
eq. \ref{eq:RPS}.

The ratios of the relativistic to nonrelativistic Fisher lengths
and impeti are presented in Fig. 1.

\section{R\'enyi entropies for the hydrogen atom}

\subsection{R\'enyi position entropies}

The R\'enyi position entropy is defined in eq. \ref{eq:Re}. The
hydrogenic nonrelativistic ground state position density yields
\begin{equation}
\label{eq:NRHr}
  H_{a}^{(r,NR)}
    =\log\left(\frac{\pi}{Z^3}\right)+3\frac{\log(a)}{a-1} \; .
\end{equation}
For $a\rightarrow 1$ this expression reduces to
$H_1^{(r,NR)}=\log\left(\frac{\pi}{Z^3}\right)+3$, which is the
well-known hydrogenic Shannon position entropy. Substituting in
eq. \ref{eq:Hs} we obtain
$${\cal{H}}_s^{(r,NR)}=\log\left(\frac{\pi}{Z^3}\right)
       +\frac{3}{2 s}\log\left(\frac{(1+s)^{1+s}}{(1-s)^{1-s}}\right) \; .$$

The hydrogenic relativistic ground state position density yields
\begin{eqnarray}
\label{eq:HrR} H_{a}^{(r,R)} &=&
\log\left(\frac{\pi\Gamma(2\gamma+1)}{2 Z^3}\right)
     +2(\gamma-1)\log(a) \\
     &+& \frac{1}{1-a} \left[ \log\left(\frac
              {\Gamma\Big( 2(\gamma-1)a +3 \Big)}
                  {\Gamma(2\gamma+1)}\right)
              -(2\gamma+1)\log(a)\right] \; . \nonumber
\end{eqnarray}
This expression is finite when the argument of the
$\Gamma$-function satisfies $2(\gamma-1)a+3>0$. This condition can
easily be traced back to the singularity of the relativistic
density at the origin, cf. eq. \ref{eq:eq11}. Since
$0\leq\gamma\leq 1$, this condition holds for all $Z$ when
$a\leq\frac{3}{2}$. For $a>\frac{3}{2}$ divergence will take place
when $Z \geq \frac{\sqrt{3(4a-3)}}{2a\alpha}$. For $\gamma=1$ eq.
\ref{eq:HrR} reduces to eq. \ref{eq:NRHr}. In the limit
$a\rightarrow 1$ we obtain the relativistic Shannon entropy, eq.
\ref{eq:position}. For $\gamma\rightarrow 0$ this expression
yields $S_r^{ER}.$

The relativistic correction $\Delta H_{a}^{(r)}=H_{a}^{(r,R)} -
H_{a}^{(r,NR)}$ can be expanded in the form
\begin{eqnarray*}
\Delta H_{a}^{(r)} &=&
\log\left(\frac{\Gamma(2\gamma+1)}{2}\right)
   +2(\gamma-1)\frac{a\log(a)}{a-1}
   -\frac{1}{a-1}\log
       \left(\frac{\Gamma(3+2\alpha(\gamma-1))}{\Gamma(2\gamma+1)}\right) \\
   &=& -\frac{a\log(a)}{a-1}\left((\alpha Z)^2 +(\alpha Z)^4/4 +\cdots\right)
       +a\left(\frac{5}{8}-\frac{\pi^2}{12}\right) (\alpha Z)^4 + \cdots
\end{eqnarray*}

Using this expansion we obtain
$$\Delta {\cal{H}}_s^{(r)}
   = -\frac{1}{2 s}\log\left(\frac{1+s}{1-s}\right)
                             \left((\alpha Z)^2 + (\alpha Z)^4/4 +\cdots\right)
    +\frac{1}{1-s^2}\left(\frac{5}{8}-\frac{\pi^2}{12}\right) (\alpha Z)^4 + \cdots$$

In the extreme relativistic limit
\begin{equation}
\label{eq:HrER} H_{a}^{(r,ER)}=\log\left(\frac{\pi
\alpha^3}{2}\right)
       +\frac{1}{1-a}\log\left(
               \frac{\Gamma(3-2a)}{a^{3-2a}}\right) \; .
\end{equation}
Eq. \ref{eq:HrER} can also be obtained from eq. \ref{eq:HrR}, by
taking the limit $\gamma\rightarrow 0$. For $a\rightarrow 1$ this
expression yields eq. \ref{eq:SER}.

The results presented above imply the commutativity of the diagram
$$\begin{array}{ccccc}
\rho_{NR}(r) & \Longrightarrow & H_{a}^{(r,NR)} &
              \longrightarrow & S_r^{NR} \\
\uparrow & & \uparrow & & \uparrow \\
\rho_R(r) & \Longrightarrow & H_{a}^{(r,R)} &
              \longrightarrow & S_r^R \\
\downarrow & & \downarrow & & \downarrow \\
\rho_{ER}(r) & \Longrightarrow & H_{a}^{(r,ER)} &
              \longrightarrow & S_r^{ER}
\end{array}$$
where
\begin{eqnarray*}
\uparrow        & & {\mbox{ stands for }}  \lim_{\gamma\rightarrow 1} \\
\downarrow      & & {\mbox{ stands for }}  \lim_{\gamma\rightarrow 0} \\
\longrightarrow & & {\mbox{ stands for }}  \lim_{a\rightarrow 1}
\end{eqnarray*}
and
$$ X(r) \Longrightarrow Y  \; {\mbox{ stands for }}
          Y=\frac{1}{1-a}\int_0^{\infty} 4\pi r^2 [X(r)]^{a} dr \, ,$$
i.e., the fact that whenever more than one path (respecting the directions
of the various arrows) is available between any two nodes, the results along
the different paths are identical.

\subsection{R\'enyi length}

Noting that $\exp(H_{a}^{(r)})$ has dimensions of volume we define
the R\'enyi length $R_{a}$ via the relation
$$\frac{4\pi}{3} R_{a}^3 = \exp(H_{a}^{(r)}) \, .$$
It follows that
$$R_{a}^{NR} = \left(\frac{3}{4}\right)^{\frac{1}{3}}
                         \frac{1}{Z}a^{\frac{1}{a-1}} \, .$$
and
$$\lim_{a\rightarrow 1} R_{\alpha}^{NR} = \left(\frac{3}{4}\right)^{\frac{1}{3}}
              \frac{e}{Z} = R_S^{NR} \, .$$
Similarly,
$$R_{a}^R = \left(\frac{3\Gamma(2\gamma+1)}{8}\right)^{\frac{1}{3}}
                 \frac{1}{Z} a^{-\frac{2a(\gamma-1)+3}{3(1-a)}}
                 \left(\frac{\Gamma(2a(\gamma-1)+3)}{\Gamma(2\gamma+1)}
                         \right)^{\frac{1}{3(1-a)}} \, ,$$
yielding $R_{a}^{NR}$ in the limit $\gamma\rightarrow 1$.

\subsection{R\'enyi momentum entropies}

The hydrogenic nonrelativistic ground state R\'enyi momentum entropy is
\begin{equation}
\label{eq:NRHp} H_{b}^{(p,NR)}
    =\log\left(\frac{\pi^2 Z^3}{8}\right)
     +\frac{1}{1-b}\log\left(I(b)\right) \; ,
\end{equation}
where,
\begin{equation}
\label{eq:I} I(b)= \frac{32}{\pi}\int_0^{\infty} (y^2+1)^{-4b}
y^2dy
  = \frac{8}{\sqrt{\pi}}\frac{\Gamma\left(4b-\frac{3}{2}\right)}
                             {\Gamma(4b)} \; .
\end{equation}
The dependence of $H_{a}^{(r,NR)}$, eq. \ref{eq:NRHr}, and
$H_{b}^{(p,NR)}$, eq. \ref{eq:NRHp}, on the nuclear charge $Z$ is
such that the sum, for any choice of $a$ and $b$, is independent
of $Z$, as demonstrated above for arbitrary homogeneous
potentials.

Noting that $I(1)=1$ we obtain, for $b\rightarrow 1$, the
nonrelativistic Shannon momentum entropy, eq. \ref{eq:eq9}.

The nonrelativistic momentum density behaves, at large $p$, as
$\Pi_{NR}(p)\sim \frac{1}{p^8}$, so that the integral in
$H_{b}^{(p,NR)}$ diverges unless $8b-2 > 1$, or $b>\frac{3}{8}$.
Indeed, for $b=\frac{3}{8}$ the numerator of eq. \ref{eq:I}
vanishes. Note, however, that this value of $b$ is below the lower
bound $b>\frac{1}{2}$ allowing the definition of the symmetrized
R\'enyi entropy, eq. \ref{eq:Hs}.

The extreme relativistic momentum density, eq. \ref{eq:eq15},
behaves, for $p\rightarrow\infty$, like
$$\Pi_{ER}(p)\sim \frac{1}{p^4} \; ,$$
so that the integral in the expression for $H_{b}^{(p,ER)}$
behaves like $\frac{1}{p^{4b-2}}$. Hence, the integral diverges
unless $4b-2>1$ or $b>\frac{3}{4}$.

The behavior of the relativistic momentum density is more subtle.
For $p\rightarrow \infty$ the variable $x$, defined in eq.
\ref{eq:x}, satisfies $x\rightarrow (\gamma+1)\frac{\pi}{2}$. As
long as $\gamma<1$ one finds that $F(p)$, defined by eq.
\ref{eq:F}, becomes a ($\gamma$ dependent) constant, so that for
large $p$ the relativistic momentum density decays as
$\Pi_R(p)\sim\frac{1}{p^{2\gamma+4}}$. It follows that the
integrand  in the expression for $H_{b}^{(p,R)}$ converges
provided that $(2\gamma+4)b-2>1$ or $b>\frac{3}{2\gamma+4}$. For
$\gamma=0$ this expression yields $b>\frac{3}{4}$, in agreement
with the result obtained above for the extreme relativistic
momentum density, but for $\gamma=1$ this expression yields
$b>\frac{1}{2}$, which is larger than the bound $b>\frac{3}{8}$
obtained above for the nonrelativistic momentum density. This is a
consequence of the fact that by taking the limit
$\gamma\rightarrow 1$ before the limit $p\rightarrow\infty$ one
obtains $F=8\frac{p^2}{(1+p^2)^2}$, that for large $p$ yields
$F\approx\frac{8}{p^2}$ rather than the constant obtained when the
limits over $\gamma$ and $p$ are taken in the opposite order.
Since $0\leq\gamma\leq 1$, it follows that for $b>\frac{3}{4}$ the
R\'enyi entropy converges for all $Z$, for $b<\frac{1}{2}$ it
diverges for all $Z$, and for $\frac{1}{2} < b < \frac{3}{4}$ it
converges for
$Z<\frac{3}{2\alpha}\sqrt{(2-\frac{1}{b})(\frac{1}{b}-\frac{2}{3}}$.
Substituting $b=\frac{1}{2-\frac{1}{a}}$ we obtain
$Z<\frac{\sqrt{3(4 a-3)}}{2a\alpha}$. Comparing with the results
obtained above for $H^{(r,R)}_a$ we conclude that $H^{(r,R)}_a$
and $H^{(p,R)}_b$ converge over the same range of $Z$ when $a$ and
$b$ are related via $\frac{1}{a}+\frac{1}{b}=2$.

The relativistic R\'enyi momentum entropies can only be obtained
numerically. The main point to note is that the various sums of
R\'enyi position and momentum entropies exhibit a dependence on Z,
unlike the nonrelativistic case. In Fig. 3 we show the sum of the
R\'enyi position and momentum entropies, $H_a^{(r)}+H_b^{(p)}$,
where $a=\frac{1}{1-s}$ and $b=\frac{1}{1+s}$, vs. $s$. The lowest
curve corresponds to the lower bound presented in eq.
\ref{eq:RUR}, and the curve just above it is the nonrelativistic
entropy sum. The relativistic entropy sums are all higher than the
nonrelativistic one, exhibiting a rapid increase for higher s
values. This behavior anticipates the approaching singularity of
the relativistic R\'enyi momentum entropy for an appropriate value
of $s>\frac{1}{3}$, that decreases with increasing $Z$, as clearly
displayed in Fig. 3. Thus, for $Z=100$ the R\'enyi momentum
entropy becomes singular for $s=\frac{2\gamma+1}{3}\approx 0.789$.

\subsection{R\'enyi impetus}

The relation
$$\frac{4\pi}{3} P_{b}^3 = \exp(H_{b}^{(p)})$$
defines the R\'enyi impetus $P_{a}$. It follows that
$$P_{b}^{NR} = \left(\frac{3\pi}{4}\right)^{\frac{1}{3}}
                       \frac{Z}{2}
          \left(\frac{8}{\sqrt{\pi}}
             \frac{\Gamma(4b-\frac{3}{2})}{\Gamma(4b)}
                 \right)^{\frac{1}{3(1-b)}}  \, .$$
In the limit $b\rightarrow 1$ this expression yields the
nonrelativistic Shannon impetus.

The relativistic R\'enyi impeti can be obtained from the
numerically evaluated R\'enyi momentum entropies.

From the uncertainty-like relation for the R\'enyi entropies, eq.
\ref{eq:RUR}, it follows that the length-impetus product satisfies
\begin{equation} \label{eq:lb}
R_aP_b\geq \left(\frac{9\pi}{16}\right)^{\frac{1}{3}}
a^{\frac{1}{2(a-1)}} b^{\frac{1}{2(b-1)}} \; .
\end{equation}
where $\frac{1}{a}+\frac{1}{b}=2$.

The ratios of the relativistic to nonrelativistic R\'enyi lengths
$(\beta_R)$ and impeti $(\mu_R)$ are presented in Fig. 2 for
conjugate pairs
$(a,b)=\left\{(\frac{5}{8},\frac{5}{2}),(\frac{3}{4},\frac{3}{2}),
(\frac{7}{8},\frac{7}{6}),(\frac{7}{6},\frac{7}{8}),(\frac{3}{2},
\frac{3}{4}),(\frac{5}{2},\frac{5}{8})\right\}$. Like the Shannon
measures presented in Fig. 1, the R\'enyi impeti show a more
pronounced relativistic effect than the corresponding lengths. The
relativistic effect on the R\'enyi impeti increases with
increasing $b$; the relativistic effect on the corresponding
lengths (that correspond to decreasing $a$, satisfying
$\frac{1}{a}+\frac{1}{b}=2$) also increases, but more moderately.

\subsection{Average position and momentum densities}

The average position density is defined as
$$<\rho>=\int_0^{\infty} 4\pi r^2 \Big(\rho(r)\Big)^2 dr \, . $$
While closely related to the 2-R\'enyi entropy, i.e.,
$<\rho>=\exp\left(-H_2^{(r)}\right)$, the average position density
merits special attention since it has recently been invoked as a
factor in a proposed measure of complexity \cite{Lopez95,Lopez02}.
The average density is also known as the Onicescu information
measure \cite{Onices01}, and is closely related to the linear
entropy $\epsilon_r = 1-<\rho>$, which is the $q=2$ case of the
Tsallis entropy \cite{Tsalis01}.

For the hydrogen atom
$$<\rho_{NR}>=\frac{Z^3}{8\pi} \, ,$$
and
$$<\rho_R>=\frac{Z^3
   \Gamma(4\gamma-1)}{\pi 2^{4\gamma-2}(\Gamma(2\gamma+1))^2} \, .$$
The relativistic expression becomes singular when $4\gamma-1=0$,
{\it{i.e.}} $Z=\frac{\sqrt{15}}{4\alpha}\approx 132.68$. The onset
of relativistic effects is given by
$\frac{<\rho_R>-<\rho_{NR}>}{Z^3} \approx 0.055159 (\alpha Z)^2
       + 0.067737 (\alpha Z)^4 +0.079452(\alpha Z)^6 +\cdots $.

The momentum density expectation value $<\Pi>=\int_0^{\infty} 4\pi
p^2 \Big(\Pi(p)\Big)^2 dp$ for the nonrelativistic density was
evaluated analytically, yielding $<\Pi_{NR}>=\frac{33}{16\pi^2
Z^3} \approx \frac{0.208975}{Z^3}$. The relativistic counterpart,
$<\Pi_R>$, can only be evaluated numerically.

The leading terms in the Taylor series expansion of
$Z^3(<\Pi_R>-<\Pi_{NR}>)$ were obtained by differentiating the
integrand, $Z^3 4\pi p^2\Big((\Pi_R(p))^2-(\Pi_{NR}(p))^2\Big)$,
with respect to $\alpha Z$, an appropriate number of times,
evaluating it at $Z=0$, and integrating over $p$. In this way we
obtain
$$Z^3(<\Pi_R>-<\Pi_{NR}>) \approx -0.254464 (\alpha Z)^2
   + 0.054446 (\alpha Z)^4 + 0.001883 (\alpha Z)^6 + \cdots \; .$$
or
$$\frac{<\Pi_R>}{<\Pi_{NR}>} \approx 1-1.21768 (\alpha Z)^2
   + 0.26054 (\alpha Z)^4 + 0.00901 (\alpha Z)^6 + \cdots \; .$$

\subsection{Average length and impetus}

Noting that $<\rho>$ has dimensions of inverse volume we define
the average length $R_A$ via
$$\frac{4\pi}{3} R_A^3 = <\rho>^{-1} \, .$$
Similarly, the average impetus $P_A$ is defined via
$$\frac{4\pi}{3} P_A^3 = <\Pi>^{-1} \, .$$
As a consequence of the connection between the average entropy and
the 2-R\'enyi entropy $H_2^{(r)}$ the lengths and impeti related
to these two information measures coincide. For the
nonrelativistic hydrogen atom we obtain
$$R_A^{NR} = \frac{6^{\frac{1}{3}}}{Z} \approx \frac{1.81712}{Z}$$
and
$$P_A^{NR} = Z \left(\frac{4\pi}{11}\right)^{\frac{1}{3}}
    \approx 1.04538 Z \, .$$
Furthermore
$$\frac{R_A^{R}}{R_A^{NR}} \approx 1-0.46210 (\alpha Z)^2
     -0.14040 (\alpha Z)^4 -0.07719 (\alpha Z)^6 + \cdots \, ,$$
and
$$\frac{P_A^{R}}{P_A^{NR}} \approx 1+0.40589 (\alpha Z)^2
     +0.24265 (\alpha Z)^4 + 0.16806 (\alpha Z)^6 +\cdots \, ,$$
hence,
$$\frac{(R_A P_A)^R}{(R_A P_A)^{NR}} \approx
        1 - 0.0562(\alpha Z)^2 + \cdots\, .$$
This expression is rather similar to the corresponding ratio for
the Fisher length and impetus, eq. \ref{eq:RPF}.

\subsection{Complexity measures}

The results presented above concerning the relativistic effects on
the various information measures for the H-like atoms allow the
evaluation of the \emph{statistical} measure of complexity $C$,
defined by L$\acute{\textrm{o}}$pez-Ruiz, Mancini, Calbet (LMC)
\cite{Lopez95,Lopez02}. The LMC measure $C$ is given by
\begin{equation}\label{eq:equ1}
   { C= H\cdot D } ~,
\end{equation}
where $H$ denotes a measure of information and $D$ represents the
so called disequilibrium or the distance from equilibrium (most
probable state). The form of $C$ is designed such that it vanishes
for the two extreme probability distributions corresponding to
perfect order $(H=0)$ and maximum disorder $(D=0)$, respectively.
It is only very recently \cite{Panos01,Panos02,Ed01,Carlos01},
that the studies on the electronic structural complexity of
\emph{neutral} atoms using the non-relativistic Hartree-Fock
$(HF)$ wave functions \cite{Bunge01} for atoms with atomic number
Z=1-54, have been reported using a variety of information
measures. A similar evaluation of a complexity measure for neutral
atoms with Z=1-103 was recently carried out in terms of the
Dirac-Fock wavefunction, choosing the exponential of the Shannon
position entropy as the measure of information, $H=\exp(S_r)$, and
the average position density as the measure of disequilibrium,
$D=<\rho>$ \cite{Borgoo}. While the measure of information
exhibits a strong shell effect but insignificant relativistic
effect, the measure of disequilibrium was found to be a
monotonically increasing function of Z exhibiting a strong
relativistic effect. It is remarkable that the ratio of the
relativistic to the nonrelativistic measures of disequilibrium
(average position densities) obtained for neutral many-electron
atoms is in almost quantitative agreement with the corresponding
ratio obtained in terms of the average position densities
evaluated above for the single electron ions. These ratios are
presented in Fig. 4. This observation must be a manifestation that
the relativistic effect on the measure of disequilibrium is
dominated by the effect on the innermost orbital.

\section{Scale invariant entropies}

\subsection{Residual position and momentum entropies}

The Kullback-Leibler relative/residual information measure of a
given probability density is defined with respect to a prior
density and it determines the extra information contained in the
given density relative to the prior. Such a residual entropy for
the relativistic density over the nonrelativistic density as the
prior can be defined in both the position and the momentum space.
In position space this can be done analytically, yielding
\begin{eqnarray}
\label{eq:eq16} S_r^{R/NR} &=& \int_0^{\infty} 4\pi r^2
       \rho_R(r)\log\left(\frac{\rho_R(r)}{\rho_{NR}(r)}\right) dr \\
   &=& \log\left(\frac{2}{\Gamma(2\gamma+1)}\right) +
     (\gamma-1)\left(2\Psi(2\gamma) + \frac{1}{\gamma}\right) \nonumber \\
  & \approx & 0.197467 (\alpha Z)^4 + 0.150105 (\alpha Z)^6
             + 0.115104 (\alpha Z)^8 + \cdots \nonumber
\end{eqnarray}

The Taylor series for $S_r^{R/NR}$ was obtained by evaluating its
first six derivatives with respect to $Z$, using maple. The values
of $S_r^{R/NR}$ for $Z<25$ were calculated using the three-term
Taylor series, since evaluating the analytic expression involves
cancellation errors. We note that for $Z=25$ the analytic
expression and the three-term expansion practically coincide.

The extreme relativistic value is obtained from eq.
(\ref{eq:eq16}) by evaluating its limit as $\gamma\rightarrow 0$.
It is found that $S_r^{R/NR}=\log(2)+2C \approx 1.847579$.

The residual momentum entropy $S_p^{R/NR} = \int_0^{\infty} 4\pi
p^2 \Pi_R(p)\log\left(\frac{\Pi_R(p)}{\Pi_{NR}(p)}\right) dp$ was
evaluated numerically. Differentiating the integrand four times
with respect to $Z$ and integrating numerically we obtained the
leading term in the power series expansion $S_p^{R/NR} \approx
0.572467 (\alpha Z)^4 + \cdots$.

The value of $\frac{S_p^{R/NR}}{(\alpha Z)^4}$ at
$Z=\frac{1}{\alpha}$ was obtained using the extreme relativistic
momentum density, eq. (\ref{eq:eq15}).

Since $S_r^{R/NR}$ and $S_p^{R/NR}$ are pure (dimensionless)
quantities, they do not measure position or momentum widths or
uncertainties. They do measure the (somewhat slow) onset of
relativistic effects upon increase of the nuclear charge.

\subsection{Average Measures of relative distance}

The average measure of relative distance \cite{Kul01,Parr01} of
the position densities is the sum of the two relative entropies
$$S_r^{R/NR}=\int_0^{\infty} 4\pi r^2 \rho_R(r)
              \log\left(\frac{\rho_R(r)}{\rho_{NR}(r)}\right) dr$$
and
$$S_r^{NR/R}=\int_0^{\infty} 4\pi r^2 \rho_{NR}(r)
              \log\left(\frac{\rho_{NR}(r)}{\rho_{R}(r)}\right) dr \; .$$
It can be written in the form
$$\tilde{S}_r=\int_0^{\infty} 4\pi r^2 \Big(\rho_R(r)-\rho_{NR}(r)\Big)
              \log\left(\frac{\rho_R(r)}{\rho_{NR}(r)}\right) dr \; .$$
The measure of relative distance of the momentum densities is
defined in an analogous manner.

$S_r^{R/NR}$ was evaluated above, {\it{cf.}} eq. (\ref{eq:eq16}).
$S_r^{NR/R}$ can be evaluated in a similar way, yielding
$$S_r^{NR/R} = \log\left(\frac{\Gamma(2\gamma+1)}{2}\right) +
(1-\gamma)(3-2C) \; .$$ Hence,
$$\tilde{S}_r=
(1-\gamma)\left(3-2C-2\Psi(2\gamma)-\frac{1}{\gamma}\right) \; .$$

The Taylor series for $\tilde{S}_r$ can be obtained analytically.
The first three terms are given by
\begin{eqnarray*}
\tilde{S}_r &\approx&
\left(\frac{\pi^2}{6}-\frac{5}{4}\right)(\alpha Z)^4 +
\left(\frac{\pi^2}{12}+\zeta(3)-\frac{7}{4}\right)(\alpha Z)^6 \\
&+&
\left(\frac{\pi^4}{90}+\frac{5\pi^2}{96}+\frac{3}{4}\zeta(3)-\frac{147}{64}\right)
        (\alpha Z)^8 +\cdots \\
&\approx& 0.394934(\alpha Z)^4 + 0.27452(\alpha Z)^6
            +0.20103(\alpha Z)^8 +\cdots
\end{eqnarray*}

$\tilde{S}_p$ was evaluated numerically.

The residual (relativistic vs. nonrelativistic) position and
momentum entropies, and the average measures of the distances of
the corresponding position and momentum distributions, are
presented in Table 1, all normalized via division by $(\alpha
Z)^4$. We note that $\frac{\tilde{S}_r}{(\alpha Z)^4}$ is a
monotonic function of $Z$, but $\frac{\tilde{S}_p}{(\alpha Z)^4}$
is not.

\section{Conclusions}

The characterization of inherent quantum mechanical uncertainties
has become a rich field of study with direct relevance to emerging
technologies. In the present article we examine the application of
widely used information measures to the ground state of the
relativistic hydrogen-like atoms, clearly bringing out the
dependence on Z due to the relativistic effects. Further, we point
out and illustrate the well-established but largely ignored
difficulties associated with the most common quantum mechanical
formulation of the uncertainty principle, that arise as a
consequence of the fact that the radial momentum is not
self-adjoint. Several information measures exhibit singularities
at particular nuclear charges, notably $Z=\frac{\sqrt{3}}{2\alpha}
\approx 118.68$ and $Z=\frac{\sqrt{15}}{4\alpha} \approx 132.68$,
whose significance remains to be elucidated. In the coordinate
representation \emph{all} the information measures considered
allowed analytic evaluation of the integrals involved. This has
not been the case for the corresponding momentum space quantities.
What we find particularly puzzling in this context is the fact
that the closed analytic expression for the position-space
expectation value of the Laplacian agrees, as expected, with the
numerically evaluated average over $p^2$, in momentum space, and
still we failed to evaluate the latter analytically. These, and
many other issues such as uncertainty and information measures for
excited states as well as for many-electron atoms, suggest that
the study of information measures for relativistic systems is a
widely open field.

\vspace{5mm}

\noindent {\bf{Acknowledgements:}} It is a great pleasure to
dedicate this work to Jesus Dehesa on the ocassion of his sixtieth
birthday. We consider it as an extension of his pioneering
contribution \cite{Yanez102} on the information entropy of
non-relativistic hydrogen atom.One of the authors (JK) wishes to
thank Professor Ady Mann for a helpful discussion. This author
also wishes to thank the School of Chemistry, University of
Hyderabad, for its hospitality.KDS acknowledges the financial
support received from the Department of Science and Technology,
New Delhi.

\vfill\eject

\vfill\eject
\newpage

\begin{table}
\caption{Residual entropies and average measures of relative
distance.} \vspace{5mm}
\begin{tabular}{|r|c|c|c|c|}
\hline
 $Z$ & $\frac{S_r^{R/NR}}{(\alpha Z)^4}$ & $\frac{S_p^{R/NR}}{(\alpha Z)^4}$ & $\frac{\tilde{S}_r}{(\alpha Z)^4}$ &
                $\frac{\tilde{S}_p}{(\alpha Z)^4}$ \\
\hline
    1  &  0.19748 &         & 0.39495 & 1.14237 \\
    2  &  0.19750 & 0.57129 & 0.39499 & 1.13967 \\
    5  &  0.19767 & 0.56733 & 0.39530 & 1.13122 \\
   10  &  0.19827 & 0.56329 & 0.39640 & 1.11703 \\
   25  &  0.20259 & 0.55744 & 0.40430 & 1.08183 \\
   50  &  0.21973 & 0.57582 & 0.43545 & 1.06601 \\
   75  &  0.25602 & 0.64370 & 0.50071 & 1.12751 \\
  100  &  0.33489 & 0.81215 & 0.63971 & 1.33467 \\
 $\frac{1}{\alpha} \;\;$  & 1.84758 & 4.03749 & 3.000000 & 5.485774 \\
\hline
\end{tabular}
\end{table}
\clearpage
\newpage

%\section{Figures}

%\begin{turnpage}
\begin{figure}
 \centering
 \includegraphics[height=10.0cm,width=10.0cm , angle=-90]{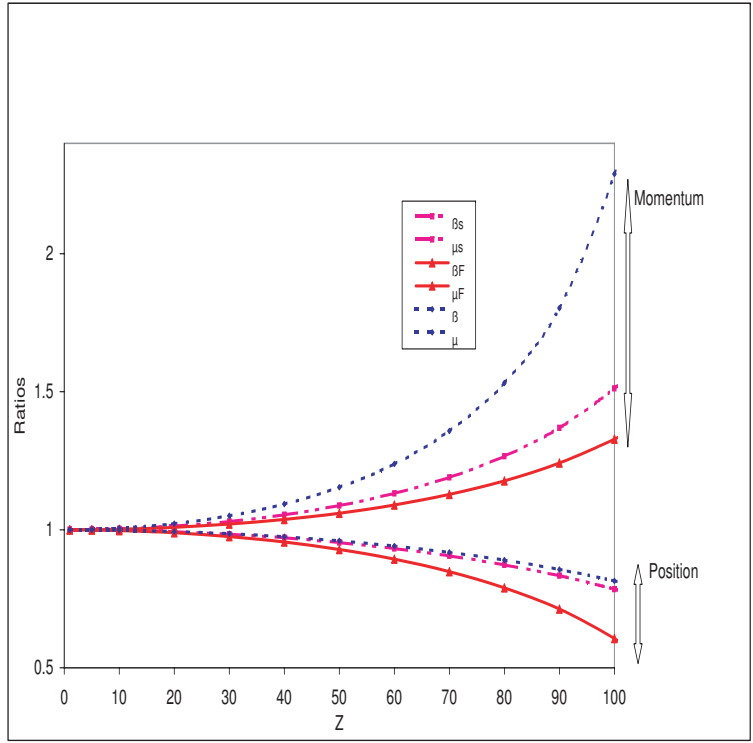}
 \caption{Ratios between the relativistic and nonrelativistic
Shannon and Fisher lengths $(\beta_S, \beta_F )$ and impeti
$(\mu_S, \mu_F )$, and corresponding ratios for root mean square
of position $(\beta )$ and momentum $(\mu )$ as functions of Z .
\label{fig:1}}
\end{figure}
%\end{turnpage}
\clearpage
\newpage

\begin{figure}
 \centering
 \includegraphics[height=10.0cm,width=10.0cm, angle=-90]{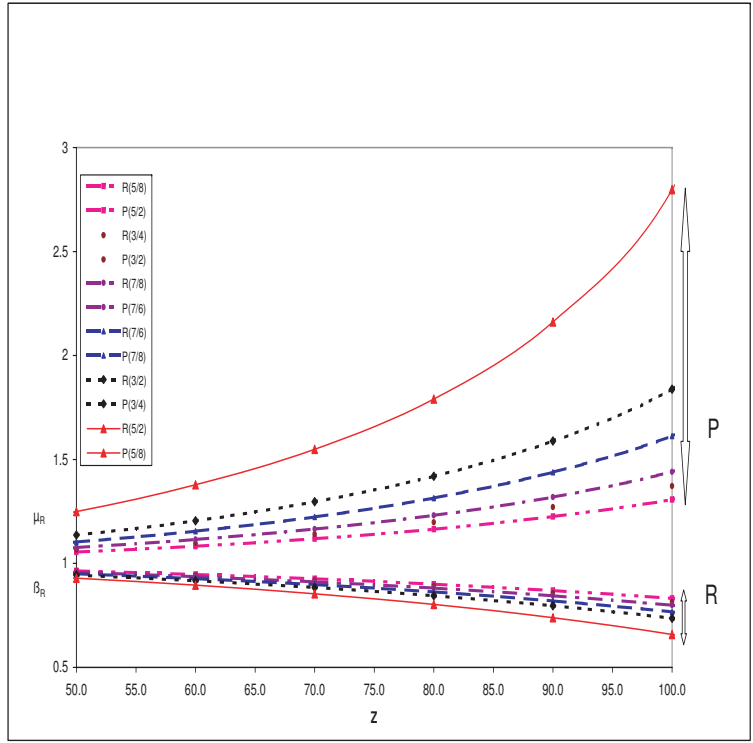}
 \caption{Ratios between the relativistic and nonrelativistic
R\'enyi lengths $(\beta_R )$ and impeti $(\mu_R )$ as functions of
Z. \label{fig:2}}
\end{figure}
\clearpage
\newpage

\begin{figure}
 \centering
 \includegraphics[height=10.0cm,width=10.0cm , angle=-90]{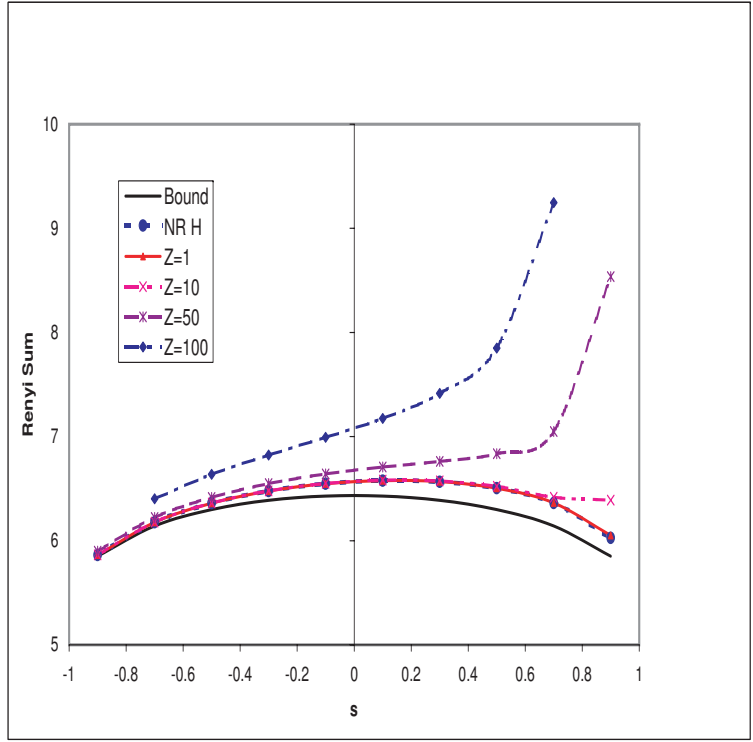}
 \caption{R\'enyi sum vs s for the non-relativistic H atom and relativistic H-like atoms. \label{fig:3}}
\end{figure}
\clearpage
\newpage

\begin{figure}
 \centering
 \includegraphics[height=10.0cm,width=10.0cm , angle=-90]{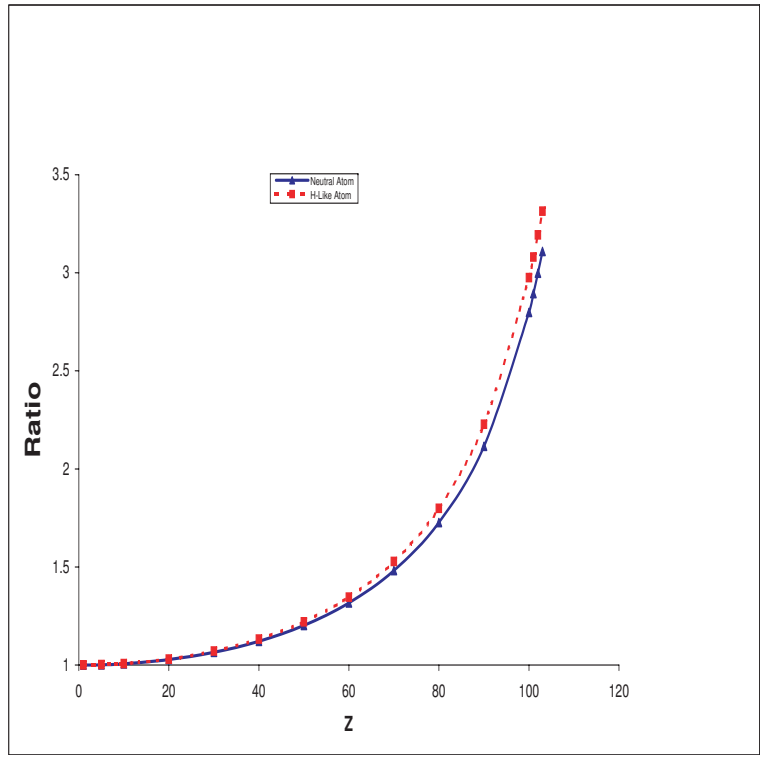}
 \caption{Ratio of relativistic to non-relativistic estimates of
the linear entropy for neutral atoms and H-like
atoms.\label{fig:4}}
\end{figure}
\clearpage

\end{document}